\documentclass[aps,prx,floatfix,twocolumn,superscriptaddress,longbibliography,nofootinbib]{revtex4-2}
\usepackage[english]{babel}
\usepackage[utf8]{inputenc}
\usepackage[T1]{fontenc}
\usepackage{physics}
\usepackage{graphicx}
\usepackage{amsmath, amsfonts, amssymb}
\usepackage{mathtools}
\usepackage[breaklinks=true, colorlinks]{hyperref}
\hypersetup{linkcolor=red, citecolor=blue}

\usepackage{bbm, bm}
\usepackage[dvipsnames]{xcolor}
\usepackage{makecell}
\usepackage{acronym}
\usepackage{upgreek}
\usepackage{nicefrac}

\def \br{\mathbf{r}}
\def \bA{\mathbf{A}}
\def \pol{\boldsymbol{\mathbf{{\varepsilon}}}}

\begin{document}

\title{Testing electron-photon exchange-correlation functional performance for many-electron systems under weak and strong light-matter coupling}

\author{Iman Ahmadabadi}
\email{imanah@umd.edu}
\affiliation{Joint Quantum Institute, NIST and University of Maryland, College Park, MD 20742, USA}
\affiliation{Center for Computational Quantum Physics, Flatiron Institute, 162 5th Avenue, New York, NY 10010, USA}

\author{I-Te Lu}
\affiliation{Max Planck Institute for the Structure and Dynamics of Matter and Center for Free-Electron Laser Science, Luruper Chaussee 149, Hamburg 22761, Germany}

\author{Leonardo A. Cunha}
\affiliation{Center for Computational Quantum Physics, Flatiron Institute, 162 5th Avenue, New York, NY 10010, USA}

\author{Michael Ruggenthaler}
\affiliation{Max Planck Institute for the Structure and Dynamics of Matter and Center for Free-Electron Laser Science, Luruper Chaussee 149, Hamburg 22761, Germany}

\author{Johannes Flick}
\affiliation{Center for Computational Quantum Physics, Flatiron Institute, 162 5th Avenue, New York, NY 10010, USA}
\affiliation{Department of Physics, City College of New York, New York, NY 10031, USA}
\affiliation{Department of Physics, The Graduate Center, City University of New York, New York, NY 10016, USA}

\author{Angel Rubio}
\email{angel.rubio@mpsd.mpg.de}
\affiliation{Max Planck Institute for the Structure and Dynamics of Matter and Center for Free-Electron Laser Science, Luruper Chaussee 149, Hamburg 22761, Germany}
\affiliation{Initiative for Computational Catalysis (ICC), The Flatiron Institute,
162 Fifth Avenue, New York, New York 10010, USA}

\begin{abstract}    
We present results of a photon-free exchange-correlation functional within the local density approximation (pxcLDA) for quantum electrodynamics density functional theory (QEDFT) that efficiently describes the electron density of many-electron systems across weak to strong light–matter coupling. Building on previous work [\href{https://journals.aps.org/pra/abstract/10.1103/PhysRevA.109.052823}{I-Te. Lu \textit{et al.}, Phys. Rev. A \textbf{109}, 052823 (2024)}] that captured electron–photon correlations via an exchange–correlation functional derived from the nonrelativistic Pauli–Fierz Hamiltonian and tested on one-electron systems, we use a simple procedure to compute a renormalization factor describing electron–photon correlations and inhomogeneity in the weak-coupling regime by comparing it with quantum electrodynamics coupled-cluster, and previous QEDFT optimized effective potential methods. Across various atoms and molecules, pxcLDA reproduces cavity-modified densities in close agreement with these references. The renormalization factor approaches unity as the system size or collective coupling increases, reflecting an electron-photon exchange-dominated behavior and improved accuracy for larger systems. This approach now offers a practical route to applying QEDFT functionals based on electron density to realistic electron systems.

\end{abstract}

\maketitle
\section{Introduction}
\label{sec:introduction}
The interaction of light and matter has emerged as a promising approach to modify material properties~\cite{ebbesen_2016a,garcia2021,ebbesen.rubio.ea_2023,bloch2022strongly}. While high-powered lasers often introduce heat~\cite{RevModPhys2021Torre, annurev2019Oka, Giovannini_2020, ahmadabadi2023optical, caruso20252025}, optical cavities can controllably modify material properties by enhancing coherence and correlations through electron–photon hybridization, forming polaritonic states~\cite{ruggenthaler.tancogne-dejean.ea_2018, flick.ruggenthaler.ea_2017, schlawin.kennes.ea_2022, kennes.rubio_2023}. These hybrid states modify ground-state reactivity and excited-state photochemical pathways, among other physical and chemical properties of materials~\cite{Thomas2019science,science2020Xiang,Anoop2016Angewandte,James2012Ange,Ahn2023Science, Bonini2024JCP,keren2025cavity}. To describe such effects, significant theoretical advances have been made recently ~\cite{jctc2025Bauman,jacsauThiam2025, JPR2023Foley,ruggenthaler.flick.ea_2014, chemrev2023Mandal, RSCS2018Ribeiro, rivera2018nanophotonic, JCP2016Kowalewski, PhysRevB2025Franco}.

Among several theoretical approaches, quantum electrodynamical density functional theory (QEDFT) has become a powerful framework to capture electron–photon interactions~\cite{tokatly_2013a, ruggenthaler.flick.ea_2014}, mainly because it incorporates the electron–photon exchange–correlation potential, which is critical for predicting cavity-modified physical properties~\cite{flick.ruggenthaler.ea_2015,pellegrini.flick.ea_2015,flick_2022}. Within QEDFT, the optimized effective potential (OEP) in the exchange approximation~\cite{pellegrini.flick.ea_2015} is accurate for single-photon processes but computationally costly. In contrast, a gradient density approximation~\cite{flick_2022} is more efficient but fails to capture the anisotropy of the cavity polarization. Both methods lose accuracy in the very strong coupling regime due to their perturbative nature~\cite{flick.schafer.ea_2018}. Recently, a photon many-body dispersion approach was developed to include anisotropy~\cite{PhysRevLett2025Tasci}. Accurate functionals remain essential for understanding cavity-induced changes to charge localization, dipole moments, and reaction dynamics~\cite{jacs2022Pavosevic,pavovsevic2023computational, Pavosevic2023jpca}. Wavefunction-based methods such as quantum electrodynamics (QED) Hartree-Fock~\cite{haugland2020PRX} and coupled-cluster (QED-CC)~\cite{PRR2020Mordovina,haugland2020PRX,Pavošević2022JCP} offer higher precision but are computationally prohibitive for large systems.

Previous work~\cite{schafer.buchholz.ea_2021,lu2024electron} introduced a non-perturbative method based on the local force equation of the non-relativistic Pauli–Fierz (PF) Hamiltonian. The recent developments of the electron-photon exchange-correlation (pxc) potential showcased promising results in simple one- and higher-dimensional one-electron systems, successfully replicating polariton spectra~\cite{schafer.buchholz.ea_2021} and electron density~\cite{lu2024electron}. In a previous study~\cite{lu2024electron}, the electron-photon exchange functional was applied to one-electron systems to capture electron–photon correlations via a renormalization factor that needs to be determined by other approaches such as exact diagonalization. However, the procedure to determine the renormalization factor has not yet been applied to three-dimensional (3D) or more complex materials.

In this work, we provide a practical procedure for obtaining the renormalization factor that captures the electron-photon correlation and inhomogeneity effects in the electron-photon exchange-correlation potential within the local density approximation (pxcLDA) for molecular 3D many-electron systems. QEDFT calculations are performed and benchmarked against QED-CC and QED optimized effective potential (QED-OEP) methods, which serve as accurate references in the weak-coupling regime~\cite{haugland2020PRX, flick.schafer.ea_2018, PRR2020Mordovina,pathak2024quantum}, while QED-CC is expected to be more reliable under strong coupling due to the perturbative limitations of OEP~\cite{pathak2024quantum, PRR2020Mordovina}. We use electron density as our benchmarking quantity, and we do not compare ground-state energies of polaritonic states here, since the photon (vector potential) information of the px(c)LDA is intentionally gauged away, leaving energy calculations for future work. This study covers a diverse set of 3D atoms and molecules that includes He, Ne, LiH, N$_2$, C$_{6}$H$_{6}$ (benzene), C$_{10}$H$_{8}$ (azulene), and sodium dimer chains (Na$_2$) and ranges from the weak to ultra-strong light–matter coupling. To improve accuracy beyond electron-photon exchange potential within the local density approximation (pxLDA), we show that it becomes crucial to introduce a renormalization factor for pxcLDA within the weak coupling regime and few electron cases, determined by minimizing a normalized squared difference between electron densities from different calculations~\cite{Bochevarov2008JCP}. Unlike pxLDA, which excludes electron-photon correlation effects, the pxcLDA functional includes them through a tuned renormalization factor. This simple approach captures missing correlation and inhomogeneity effects and reveals a general trend: as the number of electrons or effective collective coupling increases, the renormalization factor approaches unity, reflecting exchange-dominated behavior. Overall, the results show the pxcLDA approximation works well across a wide range of systems, highlighting its potential for studying cavity-induced modifications in material properties.

\section{methodology}
The PF Hamiltonian provides an exact non-relativistic first-principles description of matter coupled to quantized electromagnetic fields~\cite{haugland2020PRX,pellegrini.flick.ea_2015,flick_2022,flick.schafer.ea_2018,PhysRevLett2025Tasci,lu2024electron,schafer.buchholz.ea_2021,weber2024light,jestaedt2019AP}. Because solving the PF Hamiltonian for realistic systems is computationally prohibitive, approximate methods have been developed~\cite{ruggenthaler.sidler.ea_2023}. Within the long-wavelength approximation, we analyze three PF-based frameworks: (i) the density-based QEDFT pxcLDA functional~\cite{ruggenthaler.flick.ea_2014,schafer.buchholz.ea_2021,lu2024electron}, (ii) the orbital-dependent OEP functional of QEDFT~\cite{pellegrini.flick.ea_2015,flick.schafer.ea_2018}, and (iii) QED-CC~\cite{haugland2020PRX,Pavošević2022JCP}. This section shows the QEDFT pxc functional from the PF Hamiltonian and discusses the use of a renormalization factor for pxcLDA. We refer the interested reader to Refs.~\cite{PRR2020Mordovina,haugland2020PRX,Pavošević2022JCP,jacs2022Pavosevic} for detailed discussions of QED-CC and to Refs.~\cite{flick_2022, flick.schafer.ea_2018} for the OEP framework.

We begin with the nonrelativistic PF Hamiltonian in the long-wavelength (dipole) approximation~\cite{svendsen2025effective}. In this approximation, the PF Hamiltonian in Hartree atomic units can be written in terms of dressed photon modes~\cite{PRR2022Rokaj, lu2024electron,schafer.buchholz.ea_2021}
\begin{equation}\label{eq:H_PF_start-0}
\begin{aligned}
    \hat{H}_{\rm PF}= \hat{H}_{\rm M} + \frac{1}{c}\hat{\tilde{\mathbf{A}}}\cdot\hat{\mathbf{J}}_{p} + \hat{H}_{\gamma},
\end{aligned}
\end{equation}
with
\begin{equation}\label{eq:H_PF_start}
\begin{aligned}
    \hat{H}_{\rm M} &= -\frac{1}{2}\sum_{l=1}^{N_{e}}\nabla_{l}^{2}+\frac{1}{2}\sum_{l\neq k}^{N_{e}}w(\mathbf{r}_{l},\mathbf{r}_{k})+\sum_{l=1}^{N_{e}}v_{\rm ext}(\mathbf{r}_{l}),\\
    \hat{H}_{\gamma} &= \sum_{\alpha=1}^{M_{p}}\tilde{\omega}_{\alpha}\left(\hat{\tilde{a}}_{\alpha}^{\dagger}\hat{\tilde{a}}_{\alpha}+\frac{1}{2}\right).
\end{aligned}
\end{equation}
Here, $\hat{H}_{\rm M}$ contains the electronic kinetic energy, electron–electron interaction $w$, and external potential $v_{\rm ext}$; $\hat{H}_\gamma$ describes dressed photon modes, which include the diamagnetic contribution. Here, $N_{e}$ is the number of electrons and $M_{p}$ is the linearly polarized photon modes. The light–matter coupling term, $(1/c)\hat{\tilde{\mathbf{A}}}\cdot\hat{\mathbf{J}}_{p}$, uses the paramagnetic current operator $\hat{\mathbf{J}}_p=\sum_{l=1}^{N_e}(-i\nabla_l)$. In the long-wavelength limit, the dressed vector potential is spatially uniform,
$\hat{\tilde{\mathbf{A}}}=\sum_{\alpha=1}^{M_p}\hat{\tilde{A}}_\alpha \tilde{\boldsymbol{\varepsilon}}_\alpha$ with
$\hat{\tilde{A}}_{\alpha}=(c\tilde{\lambda}_{\alpha}/\sqrt{2\tilde{\omega}_{\alpha}})\left(\hat{\tilde{a}}^{\dagger}_{\alpha}+\hat{\tilde{a}}_{\alpha}\right)$,
where $\hat{\tilde{a}}_{\alpha}(\hat{\tilde{a}}^{\dagger}_{\alpha})$ is annihilation (creation) operator for the $\alpha$-th dressed photon mode, $\tilde{\boldsymbol{\varepsilon}}_\alpha$ is the polarization, $c$ is the light speed, $\tilde{\lambda}_{\alpha}\propto \sqrt{1/V_\alpha}$ is the mode strength (mode volume $V_\alpha$), and the dressed frequency single-mode cavity satisfies $\tilde{\omega}^{2}=\omega^{2}+N_{e}\lambda^{2}$, with $\tilde{\lambda}=\lambda$ and $\omega$ as the bare cavity frequency; weak (strong) mode strength corresponds to $\lambda/\omega\ll 1$ ($\sim 1$)~\cite{Svendsen2024JCTC,friskkockum2019NatureRev,svendsen2021combining}.

To find an approximation for many-body systems in QEDFT, we introduce a non-interacting Kohn–Sham (KS) system~\cite{lu2024electron}, which is designed to reproduce the electron density of the interacting system. We note that the ground-state problem of the Kohn–Sham system coupled to Maxwell’s equation can be solved exactly, as the zero-field condition is determined entirely by the total dipole~\cite{PRR2025Horak, sidler2025density}. The KS Hamiltonian is 
\begin{equation}\label{eq:HKS}
\begin{split}
\hat{H}_{\rm KS} &= -\frac{1}{2}\nabla^{2} + v_{\rm KS}(\mathbf{r})\\
&= -\frac{1}{2}\nabla^{2} + v_{\rm ext}(\mathbf{r}) + v_{\rm H}(\mathbf{r}) + v_{\rm xc}(\mathbf{r}) + v_{\rm pxc}(\mathbf{r}),
\end{split}
\end{equation}
where $v_{\rm H}$ is the Hartree potential, $v_{\rm xc}$ the usual electron–electron exchange–correlation term, and $v_{\rm pxc}$ the electron–photon exchange–correlation potential. 

For the electron-photon exchange-correlation potential, we approximate it with the electron-photon exchange potential within the local density approximation (LDA) for a 3D system, which satisfies the Poisson equation~\cite{lu2024electron}
\begin{equation}\label{eq:main-pxlda-d-dimension}
\nabla^{2} v_{{\rm pxLDA}}(\br) = -\sum_{\alpha=1}^{M_{p}}\frac{2 \pi^{2}\tilde{\lambda}_{\alpha}^{2}}{\tilde{\omega}_{\alpha}^{2}}\left[(\tilde{\boldsymbol{\varepsilon}}_\alpha\!\cdot\!\nabla)^{2}\left(\frac{3\rho(\br)}{8\pi}\right)^{\frac{2}{3}}\right].
\end{equation}
\begin{figure*}[!t]
  \centering
  \includegraphics[width=\linewidth]{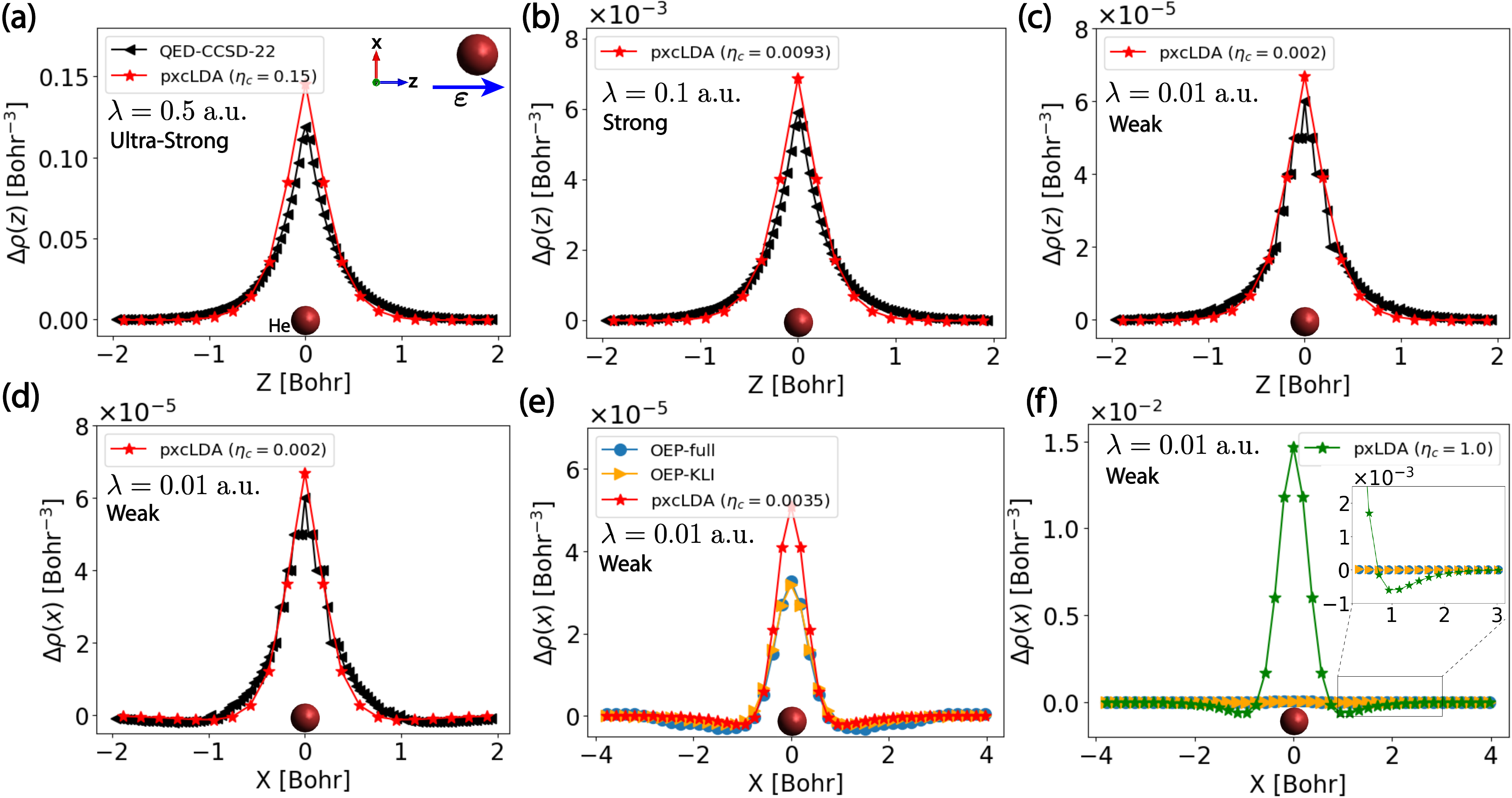}
  \caption{Calculated electron density difference ($\Delta\rho$) of the He atom between inside and outside of the cavity for various methods and physical parameters. The atom is at the center ($x=y=z=0$), and cuts pass through the nucleus (schematic red sphere). Panels (a–d) compare pxcLDA with QED-CCSD-22 at cavity frequency $\omega=2$ eV with polarization along the Z-axis for different couplings $\lambda$; (e–f) compare pxcLDA/pxLDA with OEP-full and OEP-KLI at $\lambda=0.01$ a.u. and polarization along the Z-axis.
  \label{fig:He}
}
\end{figure*}
The electron-photon exchange vanishes as $\tilde{\lambda}_{\alpha}\!\to\!0$ and dominates in the ultra-strong limit~\cite{schafer.buchholz.ea_2021}. Since the exchange-only approximation can overestimate the influence of the cavity on the electron density, it is necessary to incorporate electron-photon correlations in the weak-coupling (perturbative) regime. The perturbation analysis discussed in~\cite{lu2024electron} allows for introducing the renormalization factor $\eta_{\rm c}$ to encode electron-photon correlation for the pxLDA potential. As a result, the pxcLDA potential reads
\begin{equation}\label{eq:main-pxlda-d-dimension-corrected}
\nabla^{2} v_{{\rm pxcLDA}}(\br) = -\eta_{\rm c}\sum_{\alpha=1}^{M_{p}}\frac{2 \pi^{2}\tilde{\lambda}_{\alpha}^{2}}{\tilde{\omega}_{\alpha}^{2}}\left[(\tilde{\boldsymbol{\varepsilon}}_\alpha\!\cdot\!\nabla)^{2}\left(\frac{3\rho(\br)}{8\pi}\right)^{\frac{2}{3}}\right].
\end{equation}
Additionally, the renormalization factor $\eta_{\rm c}$ contains the $\kappa$ parameter introduced in Ref.~\cite{schafer.buchholz.ea_2021} to capture the inhomogeneity of the pxcLDA functional. The factor $\kappa$ accounts for the homogeneity of the electron system, where the maximally inhomogeneous (homogeneous) medium corresponds to $\kappa=1$ ($\kappa=0$). While the renormalization factor could be calibrated via quantum Monte Carlo simulations for the photon-coupled homogeneous electron gas, analogous to the standard LDA functional in DFT~\cite{weber2024light, Weber2025JCTC}, here we estimate $\eta_{\rm c}$ by comparing electron densities to OEP or QED-CC reference calculations. We quantify density changes induced by the cavity and compare densities across levels of theory, which can then be applied to arbitrary systems.

\section{results and discussion}

This section presents cavity-induced electron-density changes obtained with the pxcLDA for atoms and molecules, compared to QED-OEP (in exchange approximation) and QED-CC. Because solving the full OEP equation (OEP-full) is costly, we also compare to the Krieger-Li-Iafrate (KLI) approximation (OEP-KLI) in the exchange approximation~\cite{krieger.li.ea_1990,flick.schafer.ea_2018}. The pxcLDA method converges significantly faster and more efficiently than both OEP-full and OEP-KLI, since the OEP approaches rely on KS orbitals, which slow their convergence relative to pxcLDA.

To isolate cavity effects, we looked at the density difference inside versus outside the cavity, $\Delta\rho(\br)$, and quantified agreement with a reference method using the normalized squared difference~\cite{Bochevarov2008JCP},
\begin{equation}\label{I-diff}
 I = \frac{\int[\Delta\rho_{\rm Ref}(\br) - \Delta\rho_{\rm pxcLDA}(\br)]^{2}d\br}{\int[\Delta\rho_{\rm Ref}(\br)]^{2}d\br + \int[\Delta\rho_{\rm pxcLDA}(\br)]^{2}d\br}.
\end{equation}
Here, $I=0$ indicates perfect agreement, and $I=1$ corresponds to zero overlap (maximum error) density differences. The renormalization factor $\eta_{\rm c}$ is chosen to minimize $I$ for a given system and mode strength. Figure~\ref{fig:Ivsetac} (Appendix~\ref{Appendix:Suppresults}) illustrates $I(\eta_{\rm c})$ for several molecules and mode strengths versus OEP-full and OEP-KLI. The minimization is achieved by scanning around the minimum $I$ value using a step size of $0.1 \eta_{\rm c}$.

When comparing DFT-based and wavefunction-based densities, it is needed to consider that DFT pseudopotentials (PPs) remove core electrons, whereas coupled-cluster typically uses frozen cores. To avoid core-related bias, the comparisons between pxcLDA and QED-CC are performed with all-electron calculations for both~\cite{Piecuch10112010}. For QEDFT comparisons (pxcLDA vs OEP), the same PPs are used.
\begin{figure}[!t]
  \centering
  \includegraphics[width=\linewidth]{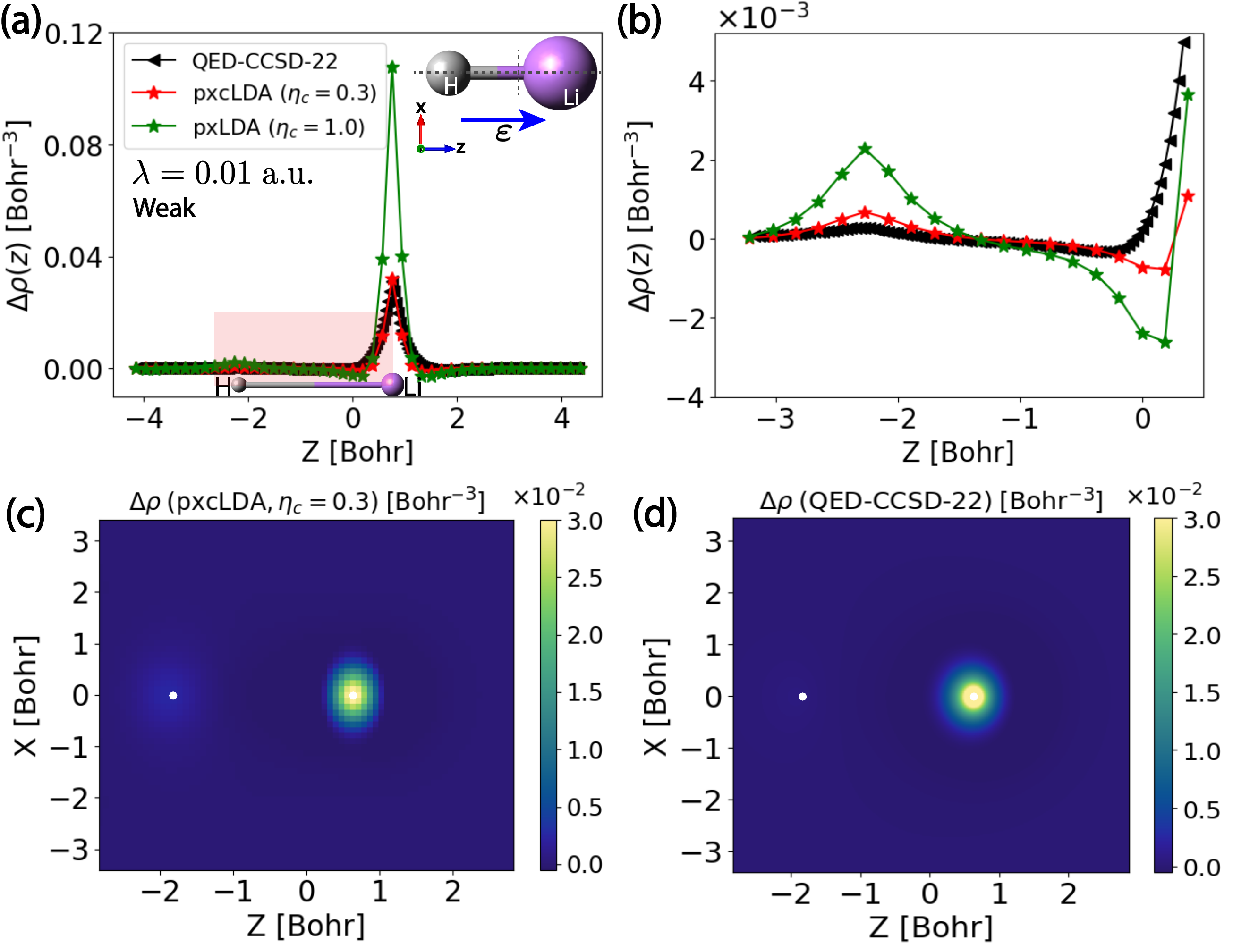}
  \caption{Electron density difference ($\Delta\rho$) of LiH between inside and outside the cavity. Panels (a–b) show pxLDA and pxcLDA compared to QED-CCSD-22 at $\omega=2$ eV, polarization along Z, with $\lambda=0.01$ a.u. and $\eta_{\rm c}=0.3$: (a) $\Delta\rho$ along the $x=y=0$ line, gray and purple spheres mark H and Li nuclei, respectively; (b) Zoomed-in views of $\Delta\rho$ at $\sim 1$ Bohr from the Li nuclei, highlighting the H–Li bond region (red area in (a)). 2D $\Delta\rho$ plots in the XZ plane ($y=0$) are shown for (c) pxcLDA with $\eta_{\rm c}=0.3$ and (d) QED-CCSD-22, with white dots marking nuclei.
  \label{fig:LiH}
}
\end{figure}
Here are the computational details for the systems we study below. For He, Ne, N$_2$, LiH, and benzene, we use the Perdew–Burke–Ernzerhof (PBE) functional~\cite{PhysRevLett1996Perdew} for electron-electron exchange-correlation with norm-conserving PBE PPs from PseudoDojo~\cite{PseudoDojo} in OCTOPUS~\cite{octopus}, a simulation radius of $11.34$ Bohr and real-space grid spacing of $0.189$ Bohr. For azulene, we use LDA~\cite{PhysRev1965KohnSham,PhysRevB1981Perdew} with Troullier–Martins PPs~\cite{PhysRevB1991Troullier}, a box of $32\times36\times16$ Bohr$^{3}$ and spacing $\Delta x=0.208$ Bohr. For sodium dimer chains, parameters and PPs follow Ref.~\cite{PhysRevB2000Kummel}; the grid is $60\times\min(60,2N_c\times10)\times60$ Bohr with 0.5 Bohr spacing, where $N_c$ is the chain length. The OEP calculations use the Barzilai-Borwein Method~\cite{IJMA1988BB, PhysRevB2012OEP} as the OEP mixing scheme. All QED-CC calculations, including those for coupled cluster with singles and doubles (CCSD) and the QED-coupled-cluster singles and doubles method up to double excitations in electronic, photonic, and mixed sectors (QED-CCSD-22), utilized the aug-cc-pVTZ basis set~\cite{Kendal1992JCP}.  The QED-CC equations are solved using the standard direct inversion of the iterative subspace method~\cite{scuseria1986CPL}, ensuring energy convergence within 10$^{-8}$ a.u for both inside and outside of the cavity simulations, using a modified implementation of the computational code presented in reference~\cite{jacs2022Pavosevic}. The following comparisons assess the accuracy, robustness, and limitations of pxcLDA relative to these references.

\subsection{He and Ne atoms in an optical cavity}

The cavity-induced electron-density difference, $\Delta\rho$, for He is shown in Fig.~\ref{fig:He}. The renormalization parameter $\eta_{\rm c}$ is obtained by minimizing $I$ in Eq.~\eqref{I-diff} between pxcLDA and either OEP-full or QED-CCSD-22. As seen in Fig.~\ref{fig:He}(a–e), pxcLDA reproduces the $\Delta\rho$ profiles along Z-axis and the X-axis from the weak ($\lambda=0.01$ a.u.) to ultra-strong ($\lambda=0.5$ a.u.) regimes, with cavity frequency $\omega=2$ eV, closely tracking QED-CCSD-22 and OEP. In contrast, pxLDA (no correlation; $\eta_{\rm c}=1$) at $\lambda=0.01$ a.u. (Fig.~\ref{fig:He}(f)) overestimates the effect; introducing $\eta_{\rm c}$ reduces the $\Delta\rho$ magnitude from the pxLDA result by $\sim 3$ orders, yielding much better agreement with the references for pxcLDA. Peak and tail positions (electron accumulation/depletion) and amplitudes are in good qualitative agreement.

The discrepancy between the pxcLDA results and the reference calculations decreases with mode strength: for instance, in He, $I_{\rm minimized}\simeq 0.248$ at $\lambda=0.01$ a.u. and $I_{\rm minimized}\simeq 0.035$ at $\lambda=0.1$ a.u. compared to OEP-full, consistent with the results for the exact electron-photon exchange functional in the infinite mode strength limit~\cite{schafer.buchholz.ea_2021}. QED-CCSD-22 and OEP-full show smaller $\Delta\rho$ at the nucleus and larger $\Delta\rho$ at distances $\sim 1$ Bohr than pxcLDA in all panels of Fig.~\ref{fig:He}. Because pxcLDA is local in the density (Eq.~\eqref{eq:main-pxlda-d-dimension-corrected})~\cite{schafer.buchholz.ea_2021,lu2024electron}, diffuse electron density regions produce weaker paramagnetic-current fluctuations and thus weaker effective local coupling, reducing the cavity’s impact there; the opposite holds near the nucleus. pxLDA overestimates accumulation at the nucleus and depletion at larger radii (inset of Fig.~\ref{fig:He}(f)), whereas pxcLDA reduces this error, since the inclusion of electron–photon correlation effects—achieved through tuned $\eta_{\rm c}$—enhances its accuracy at all positions, including both at the He nucleus and at larger radii. Ne atom inside a cavity exhibits the same behavior (Fig.~\ref{fig:Ne} in Appendix~\ref{Appendix:Suppresults}). The other examples below also show a similar behavior.

\subsection{LiH and N$_2$ in an optical cavity}

\begin{figure*}[!t]
  \centering
  \includegraphics[width=\linewidth]{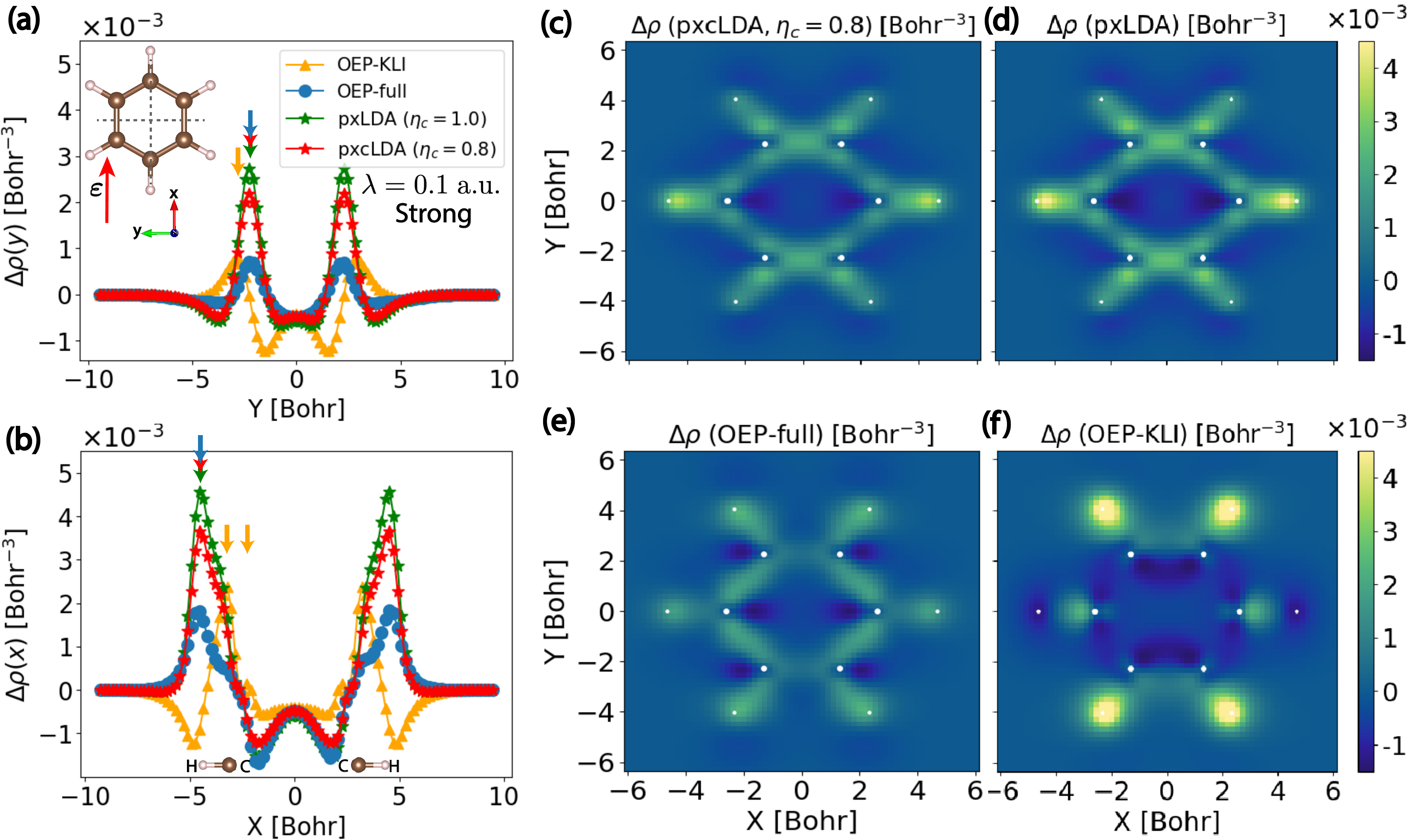}
  \caption{Electron density difference ($\Delta\rho$) of benzene inside vs. outside the cavity for OEP-full, OEP-KLI, pxLDA, and pxcLDA at $\omega=2$ eV. Panels (a–b) show 1D cuts along the Y- and X-axes for $\lambda=0.1$ a.u., highlighting that pxcLDA closely follows OEP-full, while OEP-KLI misplaces $\Delta\rho$ peaks. Panels (c–f) present 2D plots on the XY plane ($z=0$), where pxcLDA reproduces bond patterns more accurately than pxLDA and OEP-KLI; carbon and hydrogen nuclei are marked by larger and smaller white dots.
  \label{fig:Benzen}
 }
\end{figure*}

Here beyond atoms, we analyze LiH and N$_2$ with bond lengths of $3.039$ and $2.060$ Bohr, respectively. The renormalization parameter $\eta_{\rm c}$ is obtained by minimizing Eq.~\eqref{I-diff} between pxcLDA and the chosen reference.

1D diagrams of $\Delta\rho$ for LiH are plotted along the Z-axis through both nuclei in Fig.~\ref{fig:LiH}(a–b); 2D maps are shown in Fig.~\ref{fig:LiH}(c–d). The cavity polarization with mode strength $\lambda=0.01$ a.u. and cavity frequency $\omega=2$ eV is along the Li–H bond (Z-axis). Using QED-CCSD-22 as reference, pxcLDA reproduces the peak positions of $\Delta\rho$ at the H and Li nuclei (Fig.~\ref{fig:LiH}(a)) and the main features in 2D (Fig.~\ref{fig:LiH}(c–d)). pxLDA (no correlation) overestimates $\Delta\rho$ by a factor of $\sim 3$ relative to QED-CCSD-22. As in He/Ne, pxLDA predicts overly strong depletion at larger distances ($\sim 0.7$ Bohr) from Li toward +Z, whereas pxcLDA reduces this error via proper tuning of $\eta_{\rm c}$ to incorporate correlation and inhomogeneity effects. Between H and Li, pxcLDA shows slightly larger $\Delta\rho$ when moving from the left toward Li than QED-CCSD-22, and near Li, both pxLDA and pxcLDA predict greater depletion than QED-CCSD-22 (Fig.~\ref{fig:LiH}(b), red area). Since the electron density is more diffuse away from nuclei, the local coupling and cavity effect are weaker, consistent with the trends in the He and Ne atoms. Similar behavior and trend can be found in N$_{2}$ (please see Fig.~\ref{fig:N2} and its explanation in Appendix~\ref{Appendix:Suppresults}.)

\subsection{Benzene (C$_6$H$_6$) and Azulene (C$_{10}$H$_8$) in an optical cavity}

\begin{figure*}
  \centering
  \includegraphics[width=\linewidth]{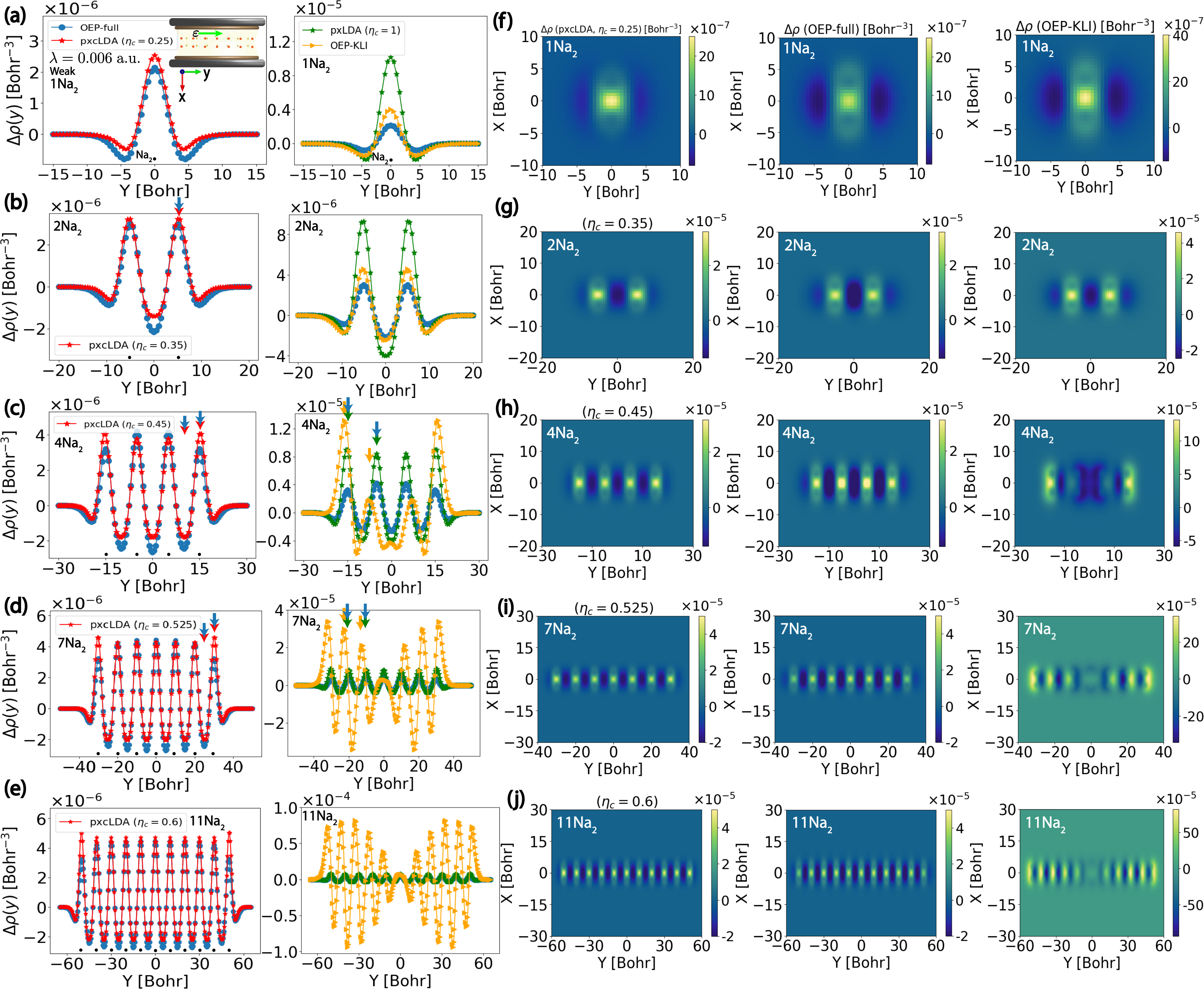}
  \caption{Electron density difference ($\Delta\rho$) between inside and outside the cavity for (a) one, (b) two, (c) four, (d) seven, and (e) eleven sodium dimers at the light-matter coupling $\lambda=0.006$ a.u., polarization along the Y-axis, and cavity frequency $\omega=2.19$ eV (resonant with the 3s–3p transition of Na$_2$). Panels (a–e, left) compare pxcLDA with OEP-full; panels (a–e, right) compare pxLDA ($\eta_{\rm c}=1$) with OEP-full and OEP-KLI. Arrows indicate minima and maxima in $\Delta\rho$, colored by functional. Black dots refer to the position of sodium dimers. 2D plots of $\Delta\rho$ for pxcLDA (left), OEP-full (middle), and OEP-KLI (right) functionals for chains with the same corresponding lengths as presented in (a-e), including (f) $N_{c}=1$, (g) $N_{c}=2$, (h) $N_{c}=4$, (i) $N_{c}=7$, and (j) $N_{c}=11$ sodium dimers in a configuration shown schematically in (a). The colorbars indicate the intensity of $\Delta \rho$ on the 2D plane. 
  \label{fig:Na2-chain-2D}
 }
\end{figure*}
The benzene and azulene molecules are also studied to assess the performance of the pxcLDA functional for larger systems with delocalized $\pi$ electrons. QED-CC calculations were not performed for benzene, since all-electron QEDFT (pxLDA and OEP) already yields highly localized densities at the nuclei. The reason is, as discussed earlier, all-electron simulations are required for correct comparison with QED-CC to avoid artifacts from PPs or frozen cores. In many-electron systems, even without electron–electron interactions, collective light–matter coupling enhances localization at nuclei with both pxLDA and OEP functionals, masking clear separation of core and valence contributions in total density in all-electron simulations. This makes the comparison with all-electron QED-CC impractical.

The benzene structure is obtained from the CCCBDB database~\cite{CCCBD} and relaxed at the second-order M{\o}ller-Plesset perturbation level of theory with the aug-cc-pVTZ basis set~\cite{Dunning1989JCP}. The cavity frequency is set to $\omega = 2.0$ eV, with mode strength $\lambda = 0.1$ a.u. and polarization aligned with the X-axis. As shown in Fig.~\ref{fig:Benzen}(a–b), pxcLDA reproduces the positions of the $\Delta\rho$ peaks (colored arrows) with higher accuracy than OEP-KLI, compared to OEP-full. Along both the X- and Y-axes, OEP-KLI deviates strongly: its peaks are shifted inward along the X-axis (to $\pm 4.5$ Bohr vs. $\pm 3.2$ Bohr for OEP-full and pxcLDA) and misaligned along the Y-axis. The pxLDA results (Fig.~\ref{fig:Benzen}(c–f)) give intensities similar to pxcLDA, indicating exchange dominates over correlation in this regime. In the 2D density plots (Fig.~\ref{fig:Benzen}(c-f)), pxcLDA reproduces both the patterns and intensities of OEP-full, including the central rhombic $\Delta\rho$ structure on the benzene ring. By contrast, OEP-KLI fails to capture the correct features (Fig.~\ref{fig:Benzen}(c–f)), missing even approximate $\Delta\rho$ between C–C bonds and failing to reproduce the rhombic pattern. Similar behavior can be observed for the azulene molecule, as another aromatic molecule, and corresponding results are presented in Fig.~\ref{fig:Azulene} in the Appendix~\ref{Appendix:Suppresults}.

\subsection{Chian of sodium dimers (Na$_2$) in an optical cavity}
As a final example, we consider chains of up to 11 sodium dimers to assess how well the pxcLDA functional captures collective light–matter interactions compared to OEP-full and OEP-KLI. QED-CC calculations are not included for the same reason as in benzene and azulene: all-electron QEDFT (pxcLDA and OEP) already shows strong nuclear localization, obscuring valence contributions and making comparisons to QED-CC impractical. Fig.~\ref{fig:Na2-chain-2D}(a-e) shows the setup and $\Delta\rho$ along the chain for various lengths ($N_c=1,2,4,7,11$), with 2D cuts shown in Fig.~\ref{fig:Na2-chain-2D}(f-j) ($z=0$ plane). The cavity frequency is tuned to the $3s$–$3p$ transition of Na$_2$ ($\omega=2.19$ eV), the mode strength set to $\lambda=0.006$ a.u., polarization aligned with the chain (Y-axis, Fig.~\ref{fig:Na2-chain-2D}(a)), and the dimer spacing fixed at $d=10$ Bohr.

The $\Delta\rho$ profiles along the Y-axis (Fig.~\ref{fig:Na2-chain-2D}(a–e)) show that pxcLDA closely follows OEP-full in both the positions and intensities of maxima and minima. As a result, electrons are pushed toward regions of higher density (the dimers), while density depletes between dimers. Consequently, robust pxcLDA and OEP-full results illustrate the same number of extrema, with pxLDA errors diminishing as $N_c$ increases (Fig.~\ref{fig:Na2-chain-2D}(a–e)). In contrast, OEP-KLI increasingly fails beyond $N_c>3$, misplacing extrema (Fig.~\ref{fig:Na2-chain-2D}(c–e) and Fig.~\ref{fig:Na2-chain-2D}(h–j)) and overestimating $\Delta\rho$ intensities by about an order of magnitude. This stems from neglect of nonlocal orbital relaxation and correlation-balance terms, causing exchange-driven overlocalization, with errors growing as the effective dipole strength increases~\cite{flick.schafer.ea_2018}.

As in He, Ne, LiH, and N$_2$, pxLDA overestimates depletion (minima) and accumulation (maxima) across the chain, but tuning the renormalization parameter $\eta_{\rm c}$ by minimizing $I$ (Eq.~\eqref{I-diff}) significantly improves pxcLDA by incorporating electron–photon correlation and inhomogeneity effects. The sodium dimer chains thus demonstrate that QEDFT can model phenomena such as the enhanced effective coupling from first principles, where many two-level systems collectively couple to the cavity.

\begin{figure}
  \centering
  \includegraphics[width=\linewidth]{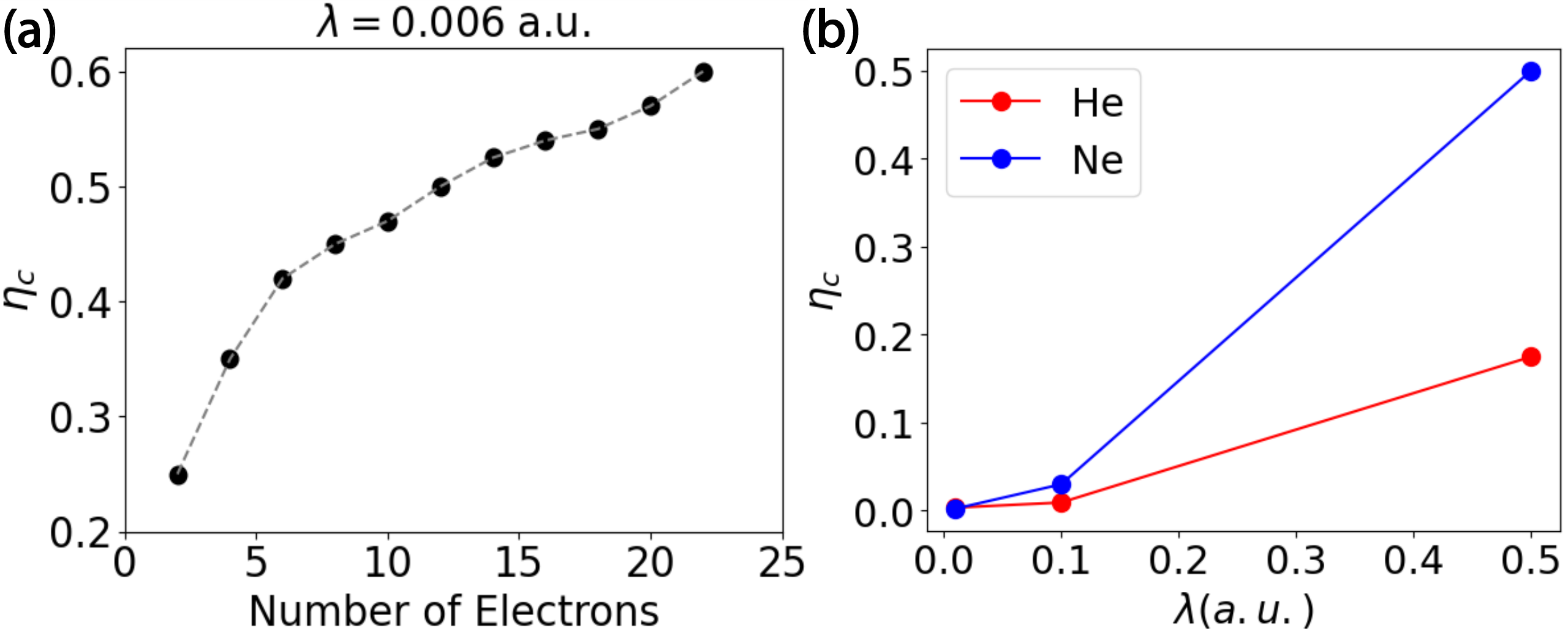}
  \caption{(a) The renormalization factor introduced in Eq.~\eqref{eq:main-pxlda-d-dimension-corrected} as a function of the number of electrons $N_{e}$ for sodium dimer chains with various lengths at $
\lambda=0.006$ a.u. (b) The renormalization factor as a function of mode strength $\lambda$ for He and Ne atoms. 
  \label{fig:etac}}
\end{figure}

So far, we have examined systems with varying electron numbers under different values of the light–matter coupling. Tracking the renormalization factor $\eta_{\rm c}$ as a function of electrons $N_{e}$ shows that $\eta_{\rm c}$ increases with $N_{e}$. This trend follows from the Breit approximation~\cite{schafer.buchholz.ea_2021}, which replaces the fluctuations of photon fields with that of matter paramagnetic current, i.e., $\Delta \hat{\tilde{\bA}} \approx -c\frac{\tilde{\lambda}_{\alpha}^{2}}{\tilde{\omega}_{\alpha}^{2}}(\tilde{\pol}_{\alpha}\cdot\Delta\hat{\mathbf{J}}_{\rm{p}})\tilde{\pol}_{\alpha}$, where $\Delta\hat{\mathbf{J}}_{\rm{p}} \propto N_{e}$. As $N_{e}$ grows, current fluctuations increase, enhancing the effective collective coupling $\lambda_{\rm eff}^{2}\propto N_{e}\tilde{\lambda}_{\alpha}^{2}/\tilde{\omega}_{\alpha}^{2}$. Since functional performance improves in the infinite mode strength limit~\cite{schafer.buchholz.ea_2021}, where exchange dominates over correlation, $\eta_{\rm c}$ increases as $N_{e}$ increases under fixed cavity conditions. Figure~\ref{fig:etac} illustrates this behavior: in sodium dimer chains $\eta_{\rm c}$ increases with $N_{e}$ (Fig.~\ref{fig:etac}(a)), reflecting collective dipole enhancement akin to the Dicke model~\cite{Garraway2011PTRS}, and also grows with increasing mode strength for a fixed system (Fig.~\ref{fig:etac}(b)). The faster increase of $\eta_{\rm c}$ for Ne compared to He indicates that, under the same simulation conditions, larger electron numbers $N_{e}$ lead to higher values of $\eta_{\rm c}$, where exchange effects dominate over electron–photon correlation effects. The results shown in Fig.~\ref{fig:etac} suggest that $\eta_{\rm c}$ can be extrapolated to larger systems.

Beyond QED-CC and OEP-full, QED auxiliary field quantum Monte Carlo provides another high-level reference~\cite{weber2024light,Weber2025JCTC}. As noted also in Refs.~\cite{lu2024electron,weber2024light}, functionals constructed solely from a strong-coupling perspective fail to generalize to weaker coupling, motivating interpolation with weak-coupling perturbation theory to include electron-photon correlations—an approach followed here. Because pxcLDA is local in density, it tends to overestimate $\Delta\rho$, as compared to accurate reference methods. Proper tuning of $\eta_{\rm c}$, however, effectively incorporates electron–photon correlations and substantially improves accuracy.
\section{Conclusion}
This work systematically evaluates the performance of the pxcLDA functional within the QEDFT framework across a broad range of systems—from atoms (He, Ne) to molecules (LiH, N$_2$, benzene, azulene) and sodium dimer chains of varying lengths in optical cavities—compared to high-accuracy QED-CC and OEP methods. We show that pxcLDA reliably captures cavity-modified electron densities in both weak and strong mode strength regimes. The functional reproduces $\Delta\rho$ peak positions near nuclei, follows OEP-full and QED-CC trends in bonding regions, and consistently outperforms OEP-KLI.

A key advance is to introduce a practical method for obtaining the renormalization
factor $\eta_{\rm c}$ for many-electron systems that can still be computed via more accurate methods such as QED-CC. While the local character of pxcLDA overestimates depletion and accumulation compared to reference methods, tuning $\eta_{\rm c}$ considerably improves accuracy over pxLDA by incorporating missing electron–photon correlation and inhomogeneity effects. This is a general procedure of minimizing the normalized squared difference (Eq.~\eqref{I-diff}) between cavity and free-space densities. We further show that $\eta_{\rm c}$ scales with the electron number $N_{e}$ and mode strength $\lambda$: as either increases, $\eta_{\rm c}$ increases, enhancing functional performance and confirming its suitability for many-electron systems and effective weak to ultra-strong coupling. Scaling laws for $\eta_{\rm c}$ to go beyond a few atom systems where reference calculations are not available will be the focus of future studies.

Future directions include correcting ground-state energies by capturing photon density energy beyond the dipole approximation, and exploring cavity-mediated solid-state materials properties and chemical reactivity~\cite{pnas2024Lu,li.cui.ea_2022,sidler.ruggenthaler.ea_2022,campos-gonzalez-angulo.poh.ea_2023,mandal.taylor.ea_2023,ruggenthaler.sidler.ea_2023,simpkins.dunkelberger.ea_2023,hirai.hutchison.ea_2023,lu2025cavity,Liu2025Optica,shin2025multiple}. Overall, this study highlights the potential of QEDFT as a framework for unifying quantum optics, chemistry, and condensed matter physics as our findings contribute to the understanding necessary for advancing applications of QEDFT in polaritonic chemistry and quantum materials science.

\section{Acknowledgment}
I.A. thanks Fabijan Pavo\v{s}evi\'{c} and Nicolas Tancogne-Dejean for helpful discussions. All calculations were performed using the computational facilities of the Flatiron Institute. The Flatiron Institute is a division of the Simons Foundation. This work was supported by the European Research Council (ERC-2024-SyG- 101167294; UnMySt). Views and opinions expressed are, however, those of the author(s) only and do not necessarily reflect those of the European Union or the European Research Council. Neither the European Union nor the European Research Council can be held responsible for them.
We acknowledge support from the Max Planck-New York City Center for
Non-Equilibrium Quantum Phenomena, the Cluster of Excellence Advanced Imaging of Matter (AIM), Grupos Consolidados (IT1249-19), and SFB925.

\appendix
\renewcommand{\thefigure}{S\arabic{figure}}
\setcounter{figure}{0}

\section{Supporting results}\label{Appendix:Suppresults}
\begin{figure}
  \centering
  \includegraphics[width=\linewidth]{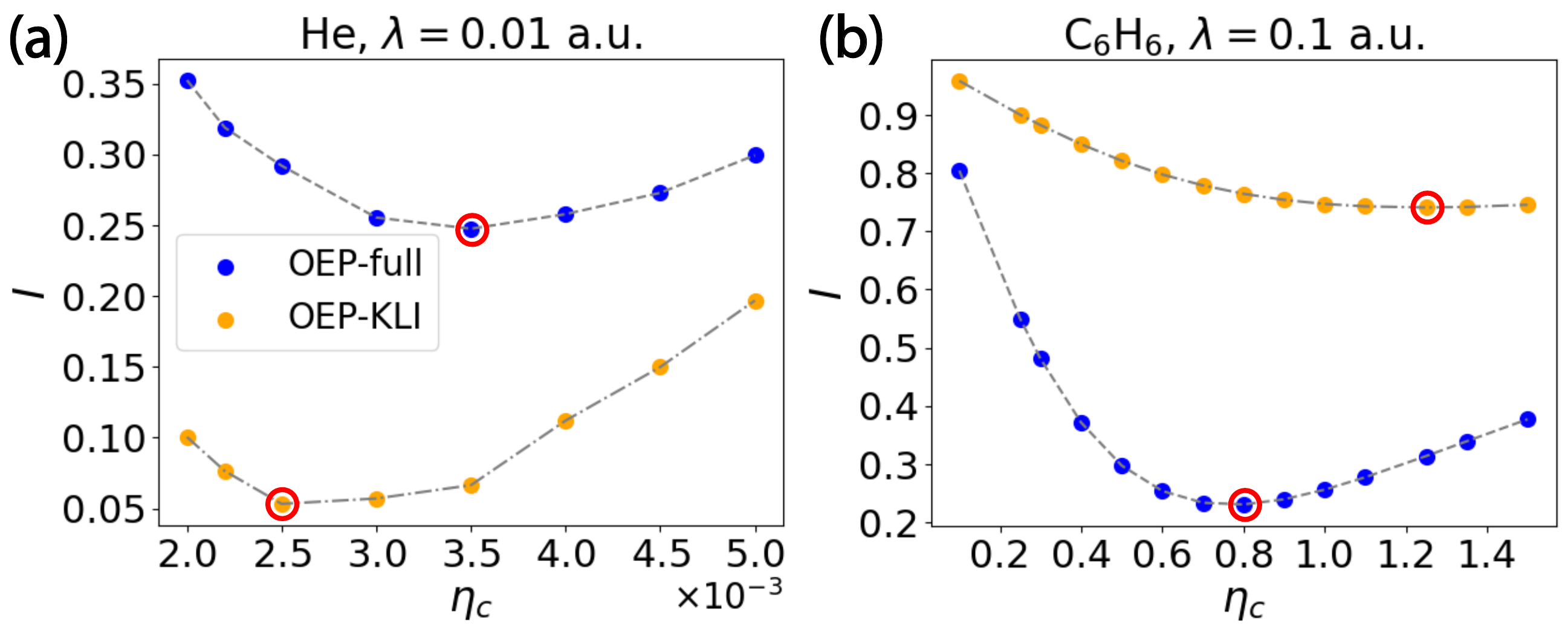}
  \caption{
  The normalized squared difference $I$ defined in Eq.~\eqref{I-diff} as a function of renormalization factor $\eta_{\rm c}$ for (a) He atom at $\lambda = 0.01$ a.u, and (b) benzene molecule at $\lambda = 0.1$ a.u. The blue and orange dots represent $I$ calculated with respect to the OEP-full and OEP-KLI as reference methods. The red circles correspond to the specific $\eta_{\rm c}$ where the minimum of $I$ happens. These results demonstrate that pxcLDA performance improves when $\eta_{\rm c}$ is properly tuned to minimize $I$. 
  \label{fig:Ivsetac}
 }
\end{figure}

The renormalization factor $\eta_{\rm{c}}$ is determined by comparing it with the results obtained using OEP or QED-CC methods, as discussed in this study. This can be done by tuning the renormalization factor $\eta_{\rm c}$ to minimize the normalized squared differences relation shown in Eq.~\eqref{I-diff}. The results for two different systems at various mode strengths are shown in Fig.~\ref{fig:Ivsetac}, indicating the minimum of normalized squared difference $I$ happening at a specific $\eta_{\rm c}$. Additionally, Fig.~\ref{fig:Ivsetac} shows that minimizing the $I$ defined in Eq.~\eqref{I-diff} with respect to the OEP-KLI results yields a different optimal value of $\eta_{\rm c}$ compared to minimization against the OEP-full results. Finally, we note that the same minimization procedure discussed in this paper can be followed for other systems.
\begin{figure*}[!t]
  \centering
  \includegraphics[width=\linewidth]{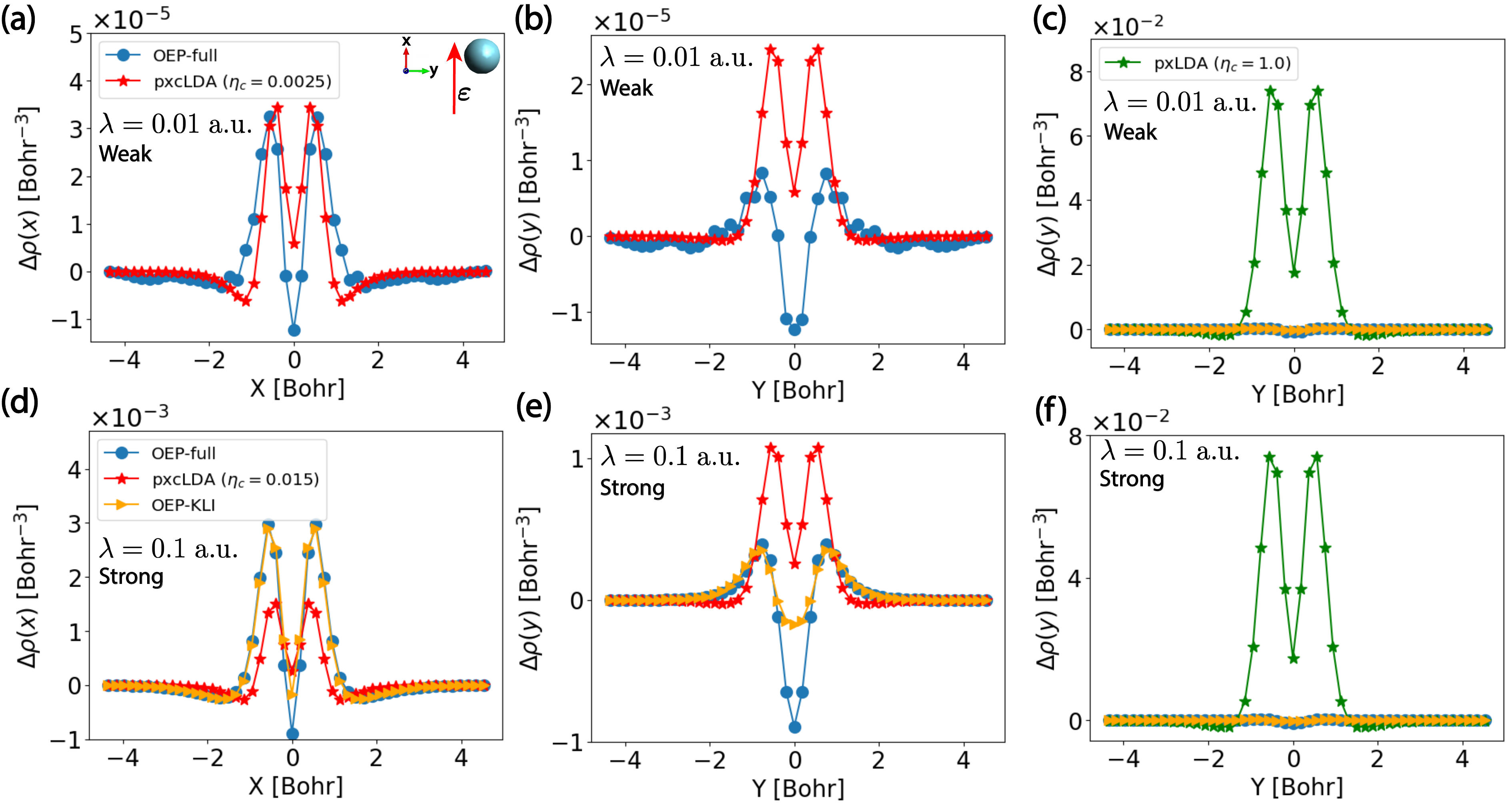}
  \caption{Electron density difference ($\Delta\rho$) of the Ne atom inside vs. outside the cavity at $\omega=2$ eV, polarization along the X-axis. Panels (a–b) show pxcLDA compared to OEP-full at $\lambda=0.01$ a.u. with $\eta_{\rm c}=0.0025$ along X- and Y-axis; panels (d–e) show the same comparison at $\lambda=0.1$ a.u. with $\eta_{\rm c}=0.015$, including the comparison with OEP-KLI. Panels (c,f) present pxLDA ($\eta_{\rm c}=1.0$) results of $\Delta\rho$ for $\lambda=0.01$ and $0.1$ a.u. along the Y-axis.
  \label{fig:Ne}
}
\end{figure*}
\begin{figure*}[!t]
  \centering
  \includegraphics[width=\linewidth]{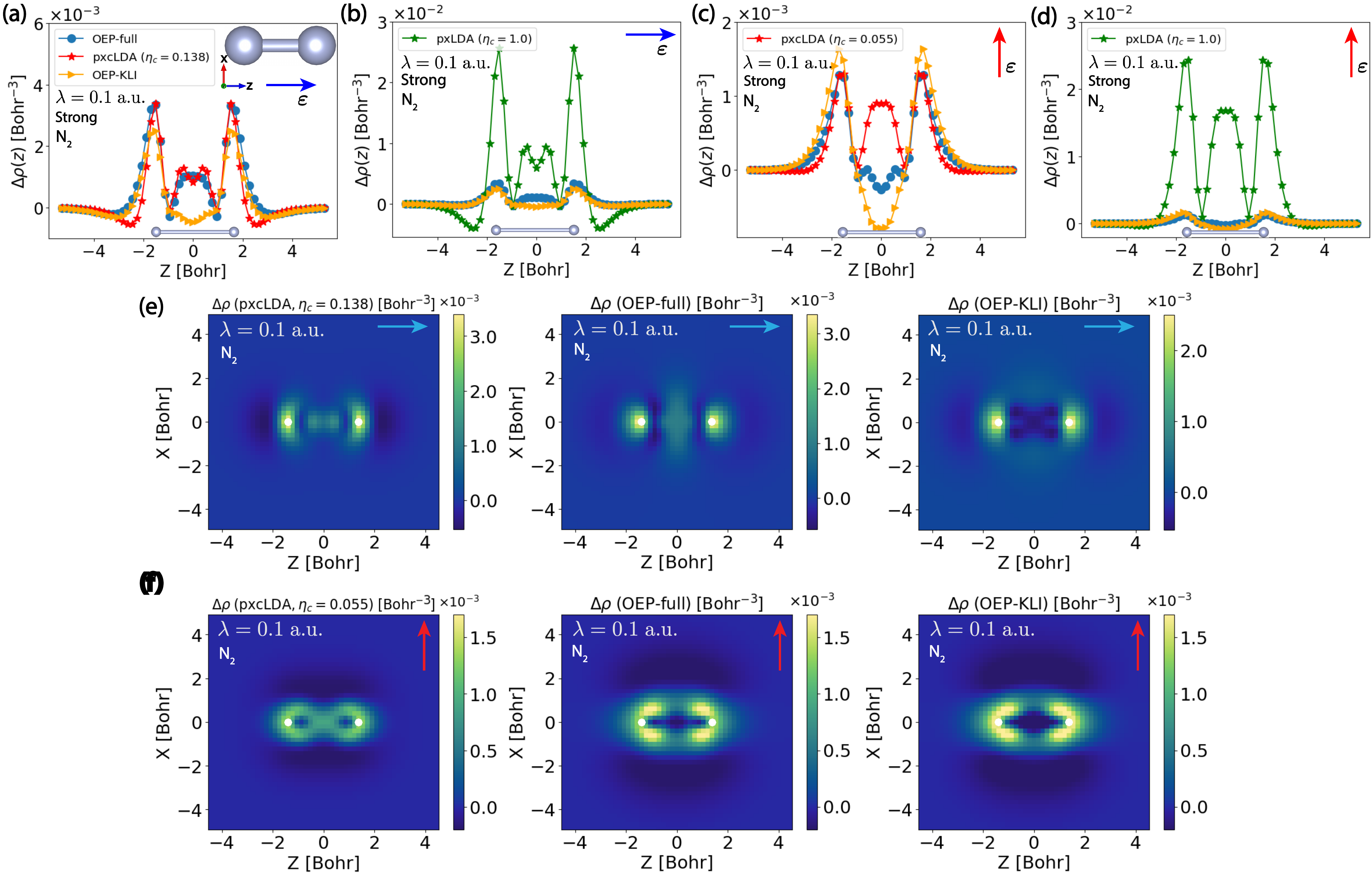}
  \caption{Electron density difference ($\Delta\rho$) of N$_2$ inside vs. outside the cavity for different functionals. Gray spheres mark nitrogen nuclei. Results are shown at $\omega=2$ eV: (a) $\lambda=0.1$ a.u., polarization along Z, pxcLDA with $\eta_{\rm c}=0.138$ compared to OEP-full and OEP-KLI; (b) pxLDA ($\eta_{\rm c}=1.0$) along Z; (c) $\lambda=0.1$ a.u., polarization along X, pxcLDA with $\eta_{\rm c}=0.055$ compared to OEP-full and OEP-KLI; and (d) pxLDA with the polarization along X. 2D $\Delta\rho$ plots on the XZ plane ($y=0$) for the same parameters are shown in (e) Z polarization and (f) X polarization, with pxcLDA (left), OEP-full (middle), and OEP-KLI (right). Small white dots mark nitrogen nuclei.
  \label{fig:N2}
}
\end{figure*}
\begin{figure}[!t]
  \centering
  \includegraphics[width=\linewidth]{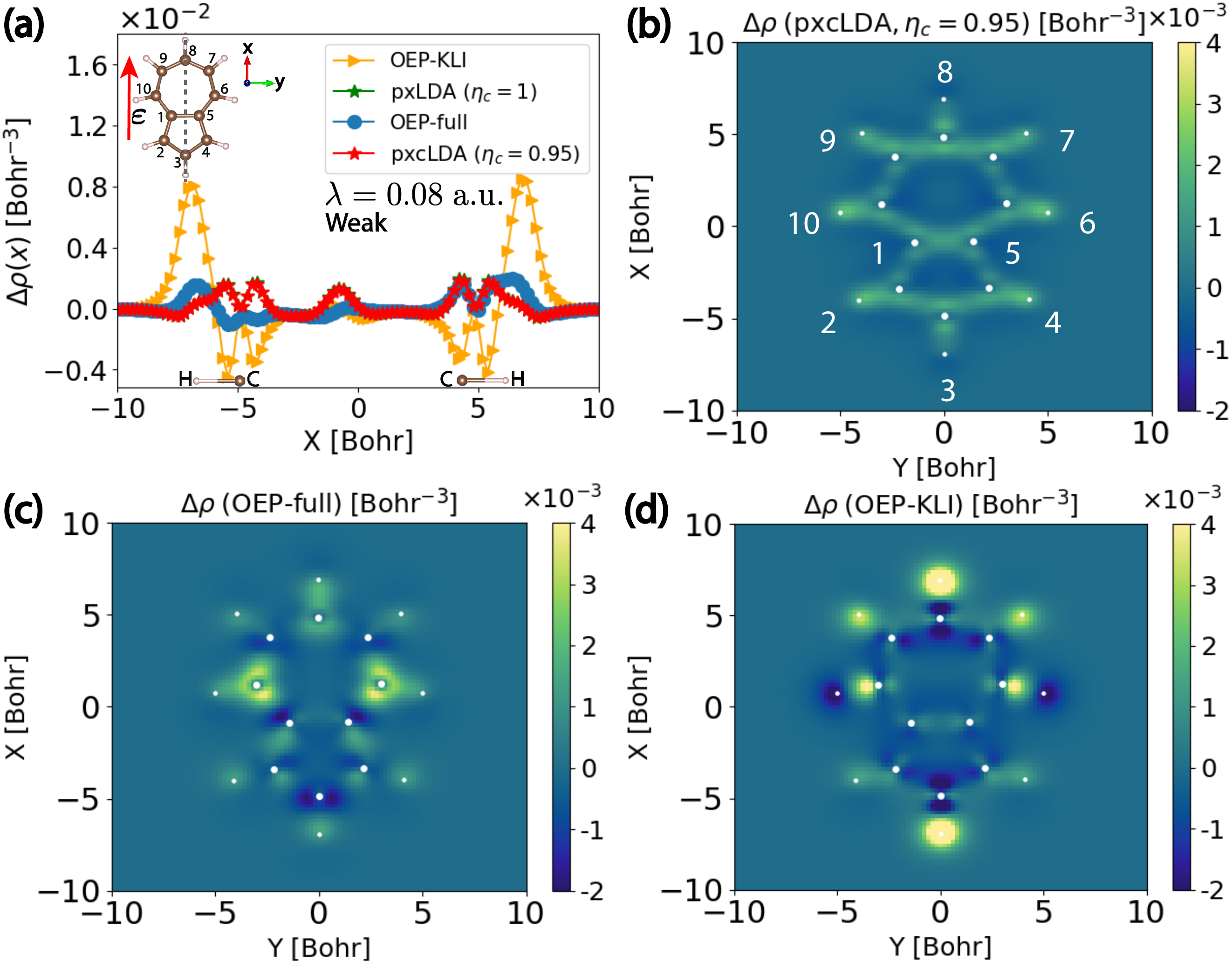}
  \caption{Electron density difference ($\Delta\rho$) of azulene between inside and outside the optical cavity for OEP-full, OEP-KLI, pxLDA, and pxcLDA functionals. Results are shown at $\omega=2.41$ eV (resonant with the HOMO–LUMO gap of the OEP-full ground state), $\lambda=0.08$ a.u., with polarization along the X-axis. (a) pxcLDA ($\eta_{\rm c}=0.95$) and pxLDA ($\eta_{\rm c}=1.0$) compared to OEP-full and OEP-KLI along the X-axis ($y=z=0$), shown as a dashed line in the schematic of azulene. Panels (b–d) show 2D $\Delta\rho$ maps on the XY plane ($z=0$) for pxcLDA, OEP-full, and OEP-KLI, with carbon and hydrogen nuclei indicated by larger and smaller white dots.
  \label{fig:Azulene}
 }
\end{figure}
This subsection provides further results for the electron density difference between inside and outside the cavity. Fig.~\ref{fig:Ne} illustrates the results of $\Delta\rho$ for the Ne atom in an optical cavity. For Ne, similar to He, $\Delta\rho$ results from the pxcLDA functional in terms of general behavior, including the regions of electron accumulation and depletion in the optical cavity, match qualitatively with the reference simulation with OEP-full functional as shown in Fig.~\ref{fig:Ne}(a-b,d-e). The results for the pxLDA and OEP-KLI are also shown in Fig.~\ref{fig:Ne}(c,f) for comparison. The error $I$ between pxcLDA and OEP-full decreases with mode strength, from $I_{\rm minimized}\simeq 0.33$ at $\lambda=0.01$ a.u. to $I_{\rm minimized}\simeq 0.29$ at $\lambda=0.1$ a.u., and pxcLDA corrects the pxLDA overestimation of $\Delta\rho$ at the nucleus and its underestimation at $\sim 1$ Bohr.

In N$_2$, as shown in Fig.~\ref{fig:N2}, the pxcLDA functional qualitatively agrees with the OEP-full results at a mode strength of $\lambda = 0.1 \text{ a.u.}$ for both polarizations that are along the Z- and X-axis, parallel and perpendicular to the N-N bond, respectively. All 1D density difference diagrams shown in Fig.~\ref{fig:N2}(a-d) are plotted on a cut passing through both N nuclei along the Z-axis. Fig.~\ref{fig:N2}(a,c) demonstrates that the pxcLDA functional can reproduce the correct peaks for $\Delta\rho$ in both polarizations where the nuclei are located, compared to the OEP-full method. The results of the pxLDA functional are shown in Fig.~\ref{fig:N2}(b,d) for comparison. Fig.~\ref{fig:N2}(e-f) shows the 2D density difference plotted on the XZ plane. Similar to He and Ne, pxcLDA predicts stronger depletion at larger distances ($\sim 1$ Bohr) from the nuclei. Its accuracy, however, is improved relative to pxLDA through electron–photon correlation and inhomogeneity effects incorporated by tuning the renormalization parameter $\eta_{\rm c}$.

For polarization along the Z-axis, pxcLDA outperforms OEP-KLI between the two N atoms, where OEP-KLI incorrectly predicts negative $\Delta\rho$ values, while both OEP-full and pxcLDA yield positive values (Fig.~\ref{fig:N2}(a,e)). For polarization along the X-axis (Fig.~\ref{fig:N2}(c,f)), pxcLDA shows overall agreement with OEP-full but with slightly reduced accuracy: it predicts electron accumulation between the N atoms, whereas OEP-full does not show significant accumulation/depletion. This is because the pxcLDA functional is local in the density, and the triple N–N bond features a highly concentrated electron distribution that couples strongly to the cavity field. The 2D density plots further demonstrate that pxcLDA captures cavity-induced density redistribution with reasonable accuracy, especially in regions not visible in the 1D profiles. These findings collectively highlight that pxcLDA provides a reliable qualitative depiction of electron density changes, although it has some limitations in certain areas depending on polarization direction.

For azulene, the 3D structure was obtained from PubChem (CID: 9231)~\cite{PubChem} and its structure relaxed with OCTOPUS with the LDA functional~\cite{PhysRev1965KohnSham,PhysRevB1981Perdew}. The cavity frequency is set to the gap between the highest occupied and lowest unoccupied molecular orbitals (HOMO-LUMO) of the OEP-full ground state of azulene ($\omega = 2.41$ eV), with mode strength $\lambda = 0.08$ a.u., polarization aligned along the X-axis, and the density differences inside and outside the cavity are shown in Fig.~\ref{fig:Azulene}. The 1D plots along the X-axis ($x=z=0$, Fig.~\ref{fig:Azulene}(a)) and 2D cuts on the molecular plane passing through carbon and hydrogen nuclei ($z=0$, Fig.~\ref{fig:Azulene}(b-d)) reveal fine structure at the molecular center and distinct density accumulation at the edges (carbons 3, 6, 8, and 10) from OEP-full results. Similar to benzene, pxLDA yields $\Delta\rho$ intensities comparable to pxcLDA, again suggesting exchange dominates correlations. As shown in Fig.~\ref{fig:Azulene}(b-d), the pxcLDA captures more detailed accumulation and depletion features around individual carbons and along C–C/C–H bonds, consistent with OEP-full, whereas OEP-KLI fails. For example, pxcLDA correctly predicts accumulation near the (6,10) carbon pair (Fig.~\ref{fig:Azulene}(b-d)), while KLI overestimates inner structures by a factor of $\sim 4$. Comparable trends are seen for pairs (1,5), (2,4), and (7,9), as well as single carbons 3 and 8. Some deviations remain in pxcLDA in comparison to OEP-full, e.g., in bonds between pairs (2,3) and (3,4) (Fig.~\ref{fig:Azulene}(b,c)). Overall, despite these limitations, pxcLDA balances accuracy and efficiency well, making it a promising functional for cavity-modified electron densities in larger molecules.
\clearpage
\bibliography{refs.bib}{}
\end{document}